\documentclass[%
reprint,
 amsmath,amssymb,
 aps,
nofootinbib
]{revtex4-1}

\usepackage{graphicx}
\usepackage{dcolumn}
\usepackage{bm}

\usepackage{amsthm}
\usepackage{color}

\definecolor{redcolor}{rgb}{0.7,0.3,0.3}
\definecolor{bluecolor}{rgb}{0.1,0.1,0.8}

\newcommand{\bfw}{\boldsymbol{w}}
\newcommand{\bfs}{\boldsymbol{s}}
\newcommand{\bfx}{\boldsymbol{x}}
\newcommand{\bfy}{\boldsymbol{y}}
\newcommand{\bfz}{\boldsymbol{z}}

\newcommand{\Real}{\mathbb{R}}
\newcommand{\Prb}{\mathbb{P}}
\newcommand{\supp}{\mathrm{supp}}

\begin{document}

\preprint{APS/123-QED}

\title{Universal principles justify the existence of concept cells}

\author{Carlos Calvo Tapia$^{1}$, Ivan Tyukin$^{2,3}$, Valeri A. Makarov$^{1,3}$}
 	\email{Corresponding author: vmakarov@ucm.es}
	\affiliation{$^1$Instituto de Matem\'atica Interdisciplinar, Faculty of Mathematics, Universidad Complutense de Madrid, Plaza Ciencias 3, 28040 Madrid, Spain\\
$^2$University of Leicester, Department of Mathematics, University Road, LE1 7RH, United Kingdom\\
$^3$Lobachevsky State University of Nizhny Novgorod, Gagarin Ave. 23, 603950 Nizhny Novgorod, Russia}

\date{\today}

\begin{abstract}
It is largely believed that complex cognitive phenomena require the perfect orchestrated collaboration of many neurons. However, this is not what converging experimental evidence suggests. Single neurons, the so-called concept cells, may be responsible for complex tasks performed by an individual. Here, starting from a few first principles, we layout physical foundations showing that concept cells are not only possible but highly likely, given that neurons work in a high dimensional space.   
\end{abstract}

\maketitle


Brain is undoubtedly high-dimensional \cite{Tyukin2018,Tozi}. Even in the simplest animal, the rotifer, it has 200 neurons acting as coupled dynamical systems, while in the human brain this figure rises to billions. The huge range of the number of neurons in different species has been related to the high variety of their cognitive abilities \cite{HH2012,MC}. 

Here, however, we assess the implication of another brain dimension, the number of synaptic inputs, $n$, a single neuron receives. Recent empirical evidence shows that a variation in the dendrite length and hence in the number of synapses $n$ can explain up to $25\%$ of the variance in IQ scores between  individuals \cite{EL}. However, no rigorous physical theory explaining  how $n$ affects high-level cognitive abilities has been put forward yet. 

The importance of such a theory and of the underlying universal principles is difficult to overestimate. For example, the design of modern artificial neural networks (ANNs) copy the converging architecture of biological sensory systems \cite{SPARSE}. As a result, they already outperform humans in pattern recognition benchmarks, yet remaining far behind in cognition \cite{ANN1,ZH}. Thus, the next qualitative leap  requires novel biophysical insights on the functional architecture of higher brain stations.

A step towards may reside in recent mathematical studies of the so-called \textit{``grandmother"} cells \cite{Tyukin2018,Gorban2018}. Converging experimental  evidence suggests that some pyramidal neurons in the medial temporal lobe can exhibit remarkable selectivity and invariance to complex stimuli. In particular, it has been shown that the so-called \textit{concept cells} (CCs) can fire when a subject sees one of 7 different pictures of Jennifer Aniston but not other 80 pictures of other persons and places \cite{QQ1}. CCs can also fire to the spoken or written name of the same person \cite{QQ2}. Thus, a single CC responds to an abstract concept but not to sensory features of the stimuli. This casts doubts on the widespread belief that complex cognitive phenomena require the perfectly orchestrated collaboration of many neurons. Moreover, CCs are relatively easily recorded in the hippocampus \cite{QQPhys}. Thus, they must be abundant in the brain, contrary to the common opinion that their existence is highly unlikely \cite{CC_Discussion}.  

Presumably, CCs play a role in episodic memory \cite{QQ2}. Memory formation and retrieval has been in the centre of attention for several decades, starting from the seminal Hopfield's work \cite{1}. Recently, the linear scaling of the memory capacity with a low factor of $0.14$ has been overcome \cite{2}. Yet, as has been found, memory retrieval is inherently unstable due to complex network dynamics \cite{3}. Thus, a ``single" neuron approach to memory functions can also be useful.

\begin{figure}[!ht]
\begin{center}
\includegraphics[width=0.48\textwidth]{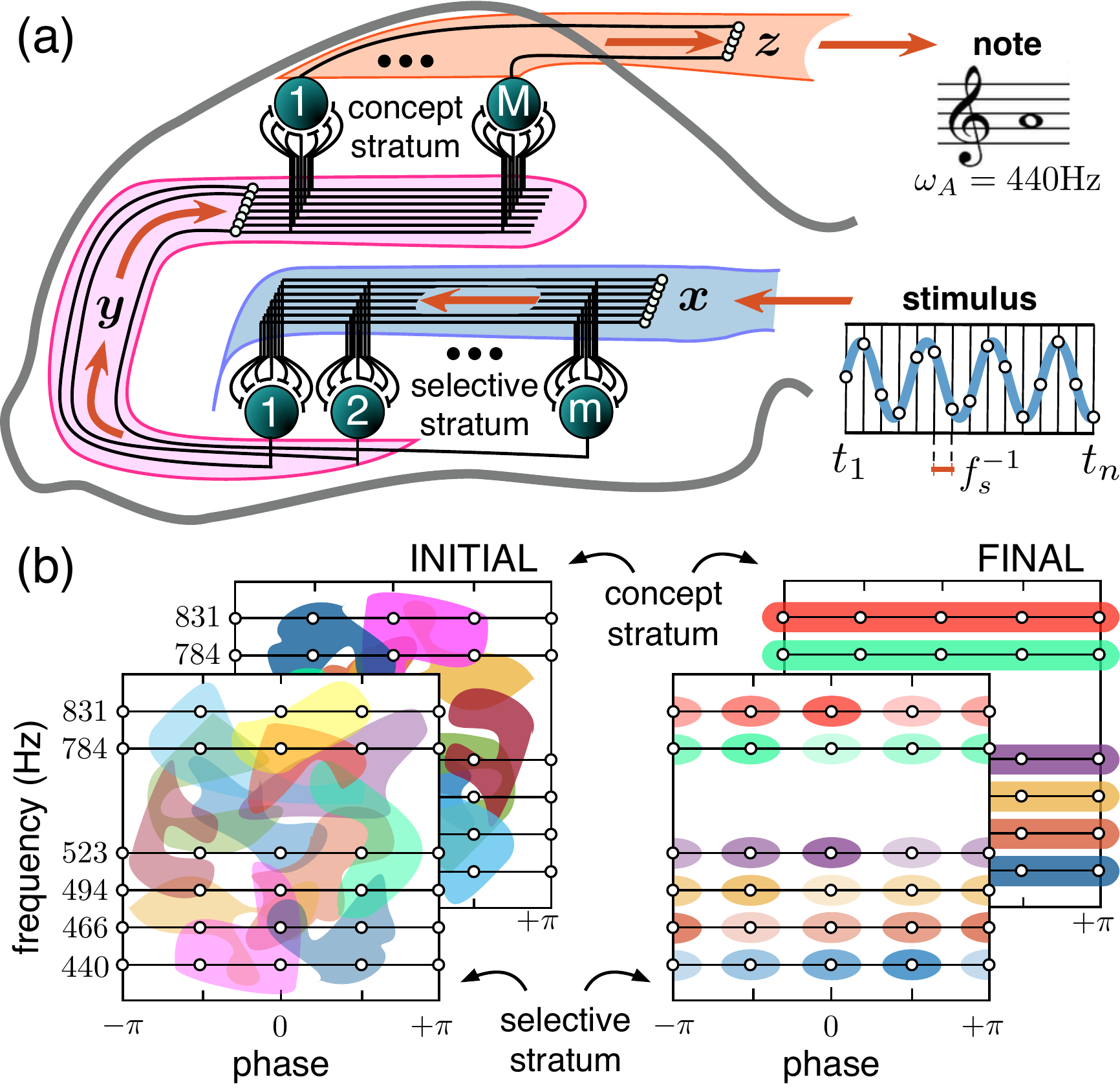}
\caption{Hypothesis of concept cells. (a) Model mimicking the information flow in the hippocampus. A stimulus (sound wave) activates the concept of a musical note. (b) Rearrangement of the neuronal ``receptive fields"  leads to formation of note-specific concept cells (different colors correspond to different neurons).
\label{Fig1}}
\end{center}
\end{figure}

In this letter, we report the first theoretical justification of the existence of CCs. It is based on a few first principles, and the neuronal dimension $n$ is the major factor. We suggest that the evolution of neurons led to an increase of their input dimension $n$, which triggered a qualitative leap in their function and emergence of episodic memory.

Figure \ref{Fig1}(a) illustrates the model mimicking basic signaling pathways in the hippocampus. It takes into account the stratified structure of the hippocampus that facilitates  ramification of axons conveying the same high-dimensional input to multiple pyramidal cells, which is supported by electrophysiological observations of spatial modules of coherent activity \cite{Oscar}. For simplicity, we consider one ``selective" and one ``concept" neural strata only, without connections between neurons within the strata.  Moreover, as we will see below, learning is localized within individual neurons. It is unsupervised and hence no global fitness function usually used in ANNs is required.  Thus, the model discards all \textit{a priori} assumptions on the local network structure and dynamics.

The first stratum contains $m$ neurons  receiving in a sequence $L$ $n$-dimensional ($n$D) stimuli $\bfx_i\in [-1,1]^n$ ($i=1,2,\ldots,L$), e.g. sound waves (Fig. \ref{Fig1}(a)). In general, $m\gg L$ (e.g. in the CA1 region of the hippocampus there are $1.4\times 10^7$ pyramidal cells). The $m$D output from the first stratum $\bfy_i$, as a response to stimulus $\bfx_i$,  goes to the second stratum. There, $K$ consecutive signals $\{\bfy_i\}$ can overlap in time due to short-term memory, implemented  through e.g. synaptic integration, and we get output $\bfz$. 

We note that neurons in the concept stratum associate several items and then respond to groups of stimuli, which form  concepts (in our case, concepts of musical notes). Stimuli within a group can be uncorrelated and even represent different sensory modalities, which gives rise to complex concepts as experimentally observed \cite{QQ2}.

Although the nature of stimuli $\bfx_i$ can be arbitrary, we illustrate this model on a simple example of acquiring ``musical memory". To follow music, the system must be able to recognize unambiguously the tones of sound or notes. To preserve generality, we do not apply any algorithmic pre-processing of sound signals, largely extended in ANNs. A piece of a sound wave sampled at $f_s=2^{13}$ Hz is represented as a ``raw" $n$D stimulus (Fig. \ref{Fig1}(a)): 
\begin{equation}
\label{Sound}
\bfx_i = \big(A_i \cos(2\pi f_i t/f_s + \phi_i)\big)_{t=1}^n, 
\end{equation}
where $A_i$, $f_i$, and $\phi_i$ are the amplitude, frequency, and phase, respectively. Let's assume that $A_i$ is fixed, i.e., the amplitude is normalized by a sensory organ. Then, $\Omega = \{(f,\phi)\}$ defines the set of primary stimuli. In this set, musical note A corresponds to frequency $440$ Hz, i.e., to the subset $\omega_A = \{(440,\phi): \forall \phi\}\subset \Omega$. 

At the beginning, all neurons in both strata are initialized randomly and hence their ``receptive fields" (areas in the sensory domain $\Omega$ invoking a response of a neuron) form a disordered mixture of random regions (see the cartoon in Fig. \ref{Fig1}(b), left). Thus, the output of the concept stratum  is random and the system cannot follow music. The purpose of learning is to organize receptive fields in such a way that the concept cells become note-specific, i.e., fire on a presentation of a given tone regardless of its phase (Fig. \ref{Fig1}(b), right). In this case, each concept cell will not be a stimulus-specific but represent a set of associated stimuli or a concept, e.g. note A.

To enable such a learning, we need at least a two-stratum system. Neurons in the sensory stratum learn to respond selectively to all sound waves, while neurons in the concept stratum associate stimuli with different phases but with the same frequency. Such an association cannot be done within the first stratum, since raw signals can be anti-correlated, e.g. $\phi_1 = 0$ and $\phi_2 = \pi$ and then cancel each other on a neuron $\bfx_1+\bfx_2 = 0$.  

We now assess the implication of the input neuronal dimensions $n$ and  $m$ on the emergence of concept cells. All neurons in both strata are described by the same model, which captures the threshold nature of the neuronal activation but disregards the dynamics of spike generation. The response  of the $j$-th neuron $y_j(t)$ in the selective stratum to the external input $\bfs_{\rm ext}(t)$ is given by:
\begin{subequations}
\label{eq:neuron_model}
\begin{align}
&\bfs_{\rm ext}= \sum_{i=1}^L\sum_{k}\sqrt{\frac{3}{n}} \bfx_i \sigma_{ik}(t),\\
&y_j=  H(v_j  - \theta_j),\ \
v_j= \langle \bfw_j,\bfs_{\rm ext} \rangle, \\
&\dot{\bfw}_j=\alpha  y_j \left(\beta^2 \bfs_{\rm ext}  -  v_j \bfw_j \right),
\end{align}
\end{subequations}
where $\sigma_{ik}(t)$ are disjoint rectangular time windows defining the $k$-th appearance of the $i$-th stimulus, $H(u)= \max\{0,u\}$ is the transfer function,  $v_j(t)$ is the membrane potential, $\theta_j \ge 0$ is the ``firing" threshold, $\bfw_j(t)\in \Real^n$ is the vector of the synaptic weights, $\langle \cdot,\cdot \rangle$ is the standard inner product, $\alpha>0$ defines the relaxation time, and $\beta>0$ is an order parameter that will be defined later. Equation (\ref{eq:neuron_model}c) simulates the Hebbian type of synaptic plasticity. The term $y_j\bfs_{\rm ext}$ forces plastic changes  when a stimulus evokes a non-zero neuronal response only, similar to the classical Oja rule \cite{Oja1982}. The second term ensures boundness of $\bfw_j$ to conform with physical plausibility.

At $t=0$, the synaptic weights of all neurons, $\bfw_j(0)$, are randomly initialized in the hypercube $U^n([-1,1])$. The threshold values $\theta_j$ can also be chosen arbitrary. Then, neuron $j$ ``fires" at the presentation of  stimulus $\bfx_i$ if its membrane potential is higher than the threshold, $v_j >\theta_j$.  In this case, we say that the neuron detects the stimulus. Let $d_j\in \{0,1,\ldots,L\}$ be the number of stimuli  the $j$-th neuron can detect.  Then, if $d_j=0$, the neuron is \emph{inactive} for $\bfs_{\rm ext}$, it is \emph{selective} if $d_j=1$, and non-selective otherwise.

To quantify the performance of the selective stratum, we introduce the ratios of selective neurons $R_{\rm slctv}$ (i.e., the number of selective neurons over $m$), inactive neurons $R_{\rm inact}$, and ``lost" stimuli $R_{\rm lost}$ (stimuli that excite no neurons). To estimate the expected values of these indexes, we note that a random  stimulus $\bfx_i$, taken from $U^n([-1,1])$, elicits random membrane potential in each neuron, which will be normally distributed as $v_j \sim \mathcal{N}(0,\frac{1}{\sqrt{3}})$, up to an error term of order $O(1/\sqrt{n})$. For $n$ large enough ($n \gtrsim 10-20$), the error decays exponentially \cite{Hoeffding1963,Suppl}. Then,  we can estimate the firing probability $\Prb(v_j > \theta) = 1 - \Phi(\sqrt{3}\theta)$ \cite{Suppl}, where $\Phi(\cdot)$ is the normal cumulative distribution function.  By using a binomial distribution, we get:
\begin{equation}
\label{meanS}
\begin{split}
\bar{R}_{\rm slctv}= &L(1-\Phi(\sqrt{3}\theta))\Phi(\sqrt{3}\theta)^{L-1},\\
\bar{R}_{\rm inact} = &\Phi(\sqrt{3}\theta)^L, \ \ \ \bar{R}_{\rm lost} = \Phi(\sqrt{3}\theta)^m.
\end{split}
\end{equation}

\begin{figure}[!ht]
\begin{center}
\includegraphics[width=0.48\textwidth]{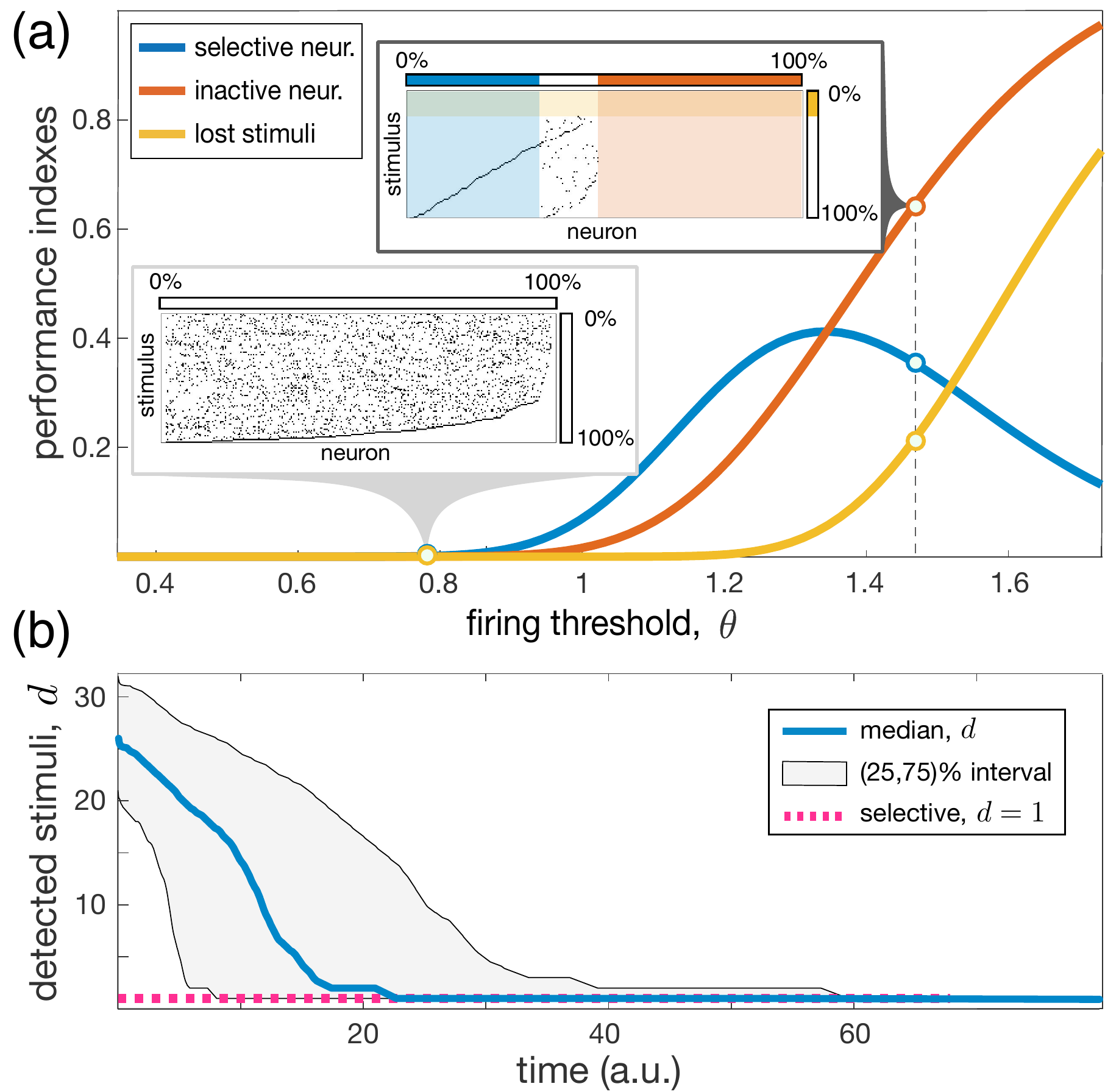}
\caption{Poor initial brain performance and the emergence of selectivity by learning. (a) Performance indexes [Eq. (\ref{meanS}), thick curves] of the selective stratum at $t=0$. Insets show raster plots of stimuli detected by neurons ($m=300$, $L = 100$).  (b) Median number of stimuli detected by a neuron, $d$, \textit{vs} time ($n = 30$, $L=400$, $\theta = 0.5$, $\alpha = 20$, $p_{\rm sl} = 0.95$).}
\label{Fig2}
\end{center}
\end{figure}

Figure \ref{Fig2}(a) illustrates the performance measures and raster plots of stimuli detected by neurons (a black dot at position $(i,j)$ means that neuron $j$ detects stimulus $i$).  
The ratio of selective neurons has a modest peak of height $e^{-1}$ at $\theta^*=\frac{1}{\sqrt{3}}\Phi^{-1}\left(\frac{L-1}{L}\right) \approx 1.35$. Therefore, at $t=0$ a randomly initialized selective stratum can have at most $37\%$ of selective neurons, independently on $n$. Thus, the first universal property of different ``brains" is their poor initial performance, regardless of the neuronal dimension $n$. 

As we show now, learning can dramatically improve the performance. We choose the firing threshold small enough, e.g. $\theta = 1$. Then, with high probability, all neurons  are active, i.e., $d_j\ge 1$, and there are no lost stimuli (Fig. \ref{Fig2}(a)), i.e., Hebbian learning (\ref{eq:neuron_model}c) is activated for all neurons and all stimuli. Figure \ref{Fig2}(b) illustrates the dynamics of the median number of the stimuli detected by neurons. At $t=0$, all neurons in the aggregate are not selective and respond in average to $d=25$ stimuli, while at $t=80$ they  are absolutely selective, $d= 1$. 

To extend this numerical observation, we first find the condition that a neuron, started firing to a stimulus $\bfx_i$, keeps firing in forward time with a probability no smaller than some constant $0<p_{\rm sl}<1$. This condition is fulfilled by choosing the order parameter  \cite{Suppl}:  
\begin{equation}
\label{SelBeta}
\beta_{\rm sl} =  \frac{\theta}{\delta}, \ \ \delta = \sqrt{1-\frac{2\Phi^{-1}(p_{\rm sl})}{\sqrt{5n}}}.
\end{equation}
Note that the higher  the neuronal dimension $n$, the higher $p_{\rm sl}$ can be chosen. 

Under condition (\ref{SelBeta}) the synaptic weights converge \cite{Suppl}:
\begin{equation}
\label{winf1}
\lim_{t\to \infty} \bfw(t) = \beta_{\rm sl}\frac{\bfx_i}{\|\bfx_i\|}.
\end{equation}
Thus, given that $\alpha$ is large enough, learning forces the neuron to ``align'' along its ``preferable'' stimulus: $\bfw_\infty \upuparrows \bfx_i$. At the same time, for high $n$, we have the property: $\langle \bfx_i,\bfx_k\rangle \approx 0$, $i\ne k$ (for details see e.g. \cite{Gorban2018}). Thus, after a transient, the neuronal membrane potential will be close to zero for all stimuli except $\bfx_i$ and hence the neuron will become selective. 

To estimate the probability that a neuron will be selective after learning, i.e. $S:=\Prb (d_j = 1)$, we evaluate the probability that the neuron will be silent to another arbitrary stimulus $\bfx_k$ ($k\ne i$) \cite{Suppl}: 
\begin{equation}
\label{PSilent1}
\mathcal{P} = \int_0^\infty \Phi(\delta\sqrt{ns}) \kappa(s;\mu,\sigma)\, ds,
\end{equation}
where $\kappa(\cdot;\mu,\sigma)$ is the normal pdf with the mean $\mu = 1$ and the standard deviation $\sigma = \frac{2}{\sqrt{5n}}$. Note that with an increase of $n$, $\kappa$ concentrates around $1$ and we can roughly evaluate $\mathcal{P}\approx \Phi(\delta\sqrt{n})$, which tends to $1$ for high $n$.  
Finally, the neuronal selectivity,
\begin{equation}
\label{PSilent2}
S(n,L)  = \mathcal{P}^{L-1},
\end{equation}
depends on the number of stimuli $L$ and the neuronal dimension $n$ only. 

\begin{figure}[!hbt]
\begin{center}
\includegraphics[width=0.49\textwidth]{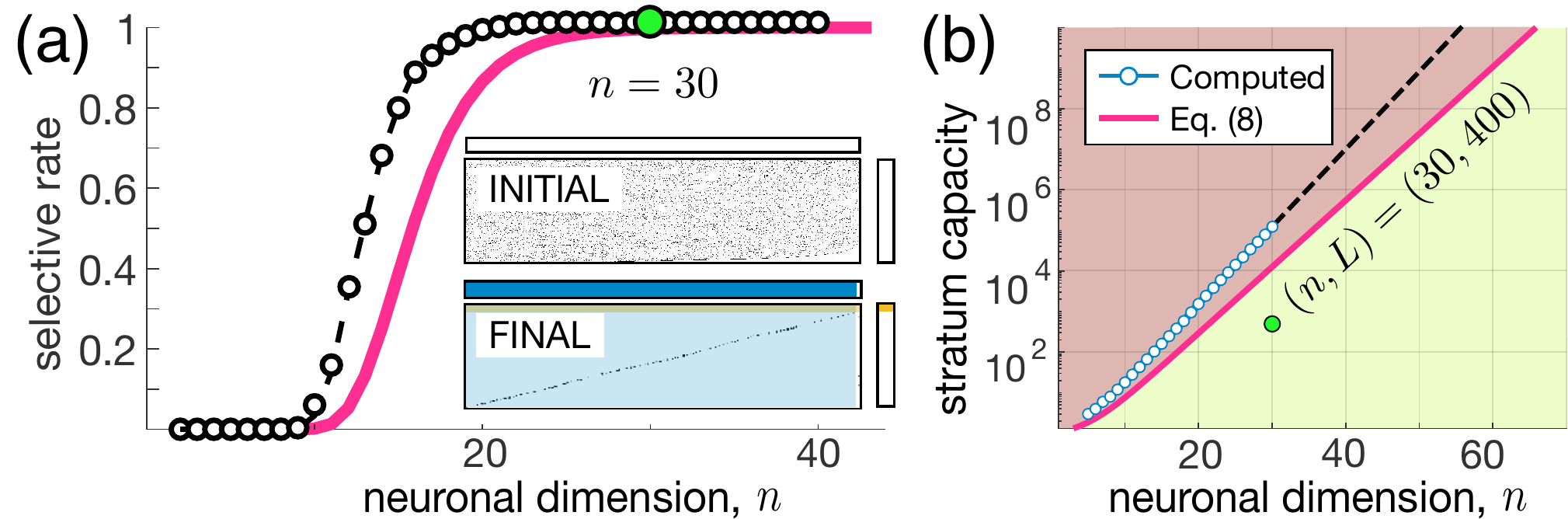}
\caption{Emergence of extreme selectivity in high-dimensional brains.  (a) Step-like increase of the ratio of selective neurons (experiment: black circles, estimate (\ref{PSilent2}): red curve). Insets show paster plots of detected stimuli for $n=30$ (compare to Fig. \ref{Fig2}(a)). (b) Exponential growth of the memory capacity (experiment: blue circles, theoretical estimate: red curve). Green area marks the working zone. }
\label{Fig3}
\end{center}
\end{figure}

Figure \ref{Fig3}(a) shows the selectivity $S$ as a function of the neuronal dimension $n$. Learning yields a step-like dependence of $S$ on $n$. For  small $n$, there is no any improvement of the selectivity by learning ($S\approx 0$), while for higher $n$, it rapidly reaches 100\%. Insets in Fig. \ref{Fig3}(a) illustrate an example of raster plots of stimuli detected by the neurons at the beginning and the end of learning. We observe that almost all neurons become selective to single information items (area shadowed by blue). Thus, the second universal property is that an increase in the input dimension $n$ provokes an explosive emergence of selective behavior in ``brains" composed of high-dimensional neurons at critical dimensions 10--20.

From Eq. (\ref{PSilent2}) we can also estimate the maximal  number of stimuli that a big enough stratum can work with:
\begin{equation}
\label{EstimM}
L_{\max} =  1 + \frac{\ln(p_L)}{\ln(\mathcal{P}) },
\end{equation}
where $p_L$ is the lower bound of the probability that the stratum detects all $L$ stimuli. Figure \ref{Fig3}(b) shows the theoretical and experimental estimates of the stratum  capacity. Even for a rather moderate dimension $n=60$, the capacity goes beyond $10^{10}$ (numerical estimate beyond  $n=30$ was not calculated due to exponential growth of the computational load). In practical terms, it means that a big enough ``brain" consisting of high-dimensional neurons can selectively detect all stimuli existing in the world. 

\begin{figure}[!hbt]
\begin{center}
\includegraphics[width=0.49\textwidth]{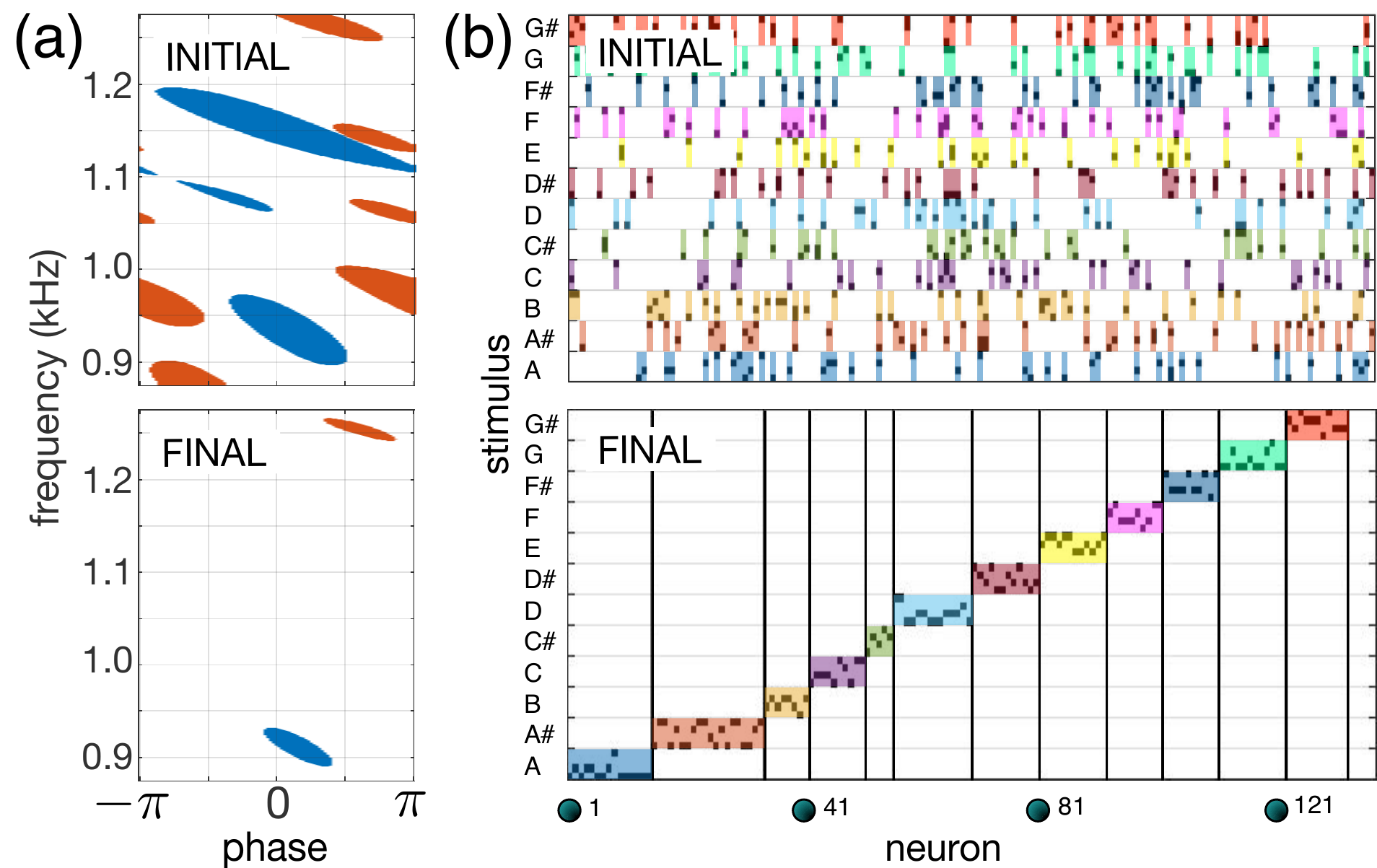}
\caption{Learning musical stimuli.  (a) Receptive fields of two neurons in the ``sound" space before and after learning. (b) Raster plot at $t=0$ (top) shows random response of the stratum to 48 stimuli representing 12 musical notes from A to G\#. After learning (bottom),  neurons have grouped into clusters selective to individual stimuli.}
\label{Fig3b}
\end{center}
\end{figure}

To illustrate how the selective stratum can deal with ``real-world" stimuli, we simulated learning of 48 sound waves  corresponding to 12 musical notes from A to G\# [see Eq. (\ref{Sound}) and Fig. \ref{Fig1}].  Figure \ref{Fig3b}(a) shows the receptive fields of two arbitrary chosen neurons before and after learning. At the beginning, the neurons have wide random receptive fields, as it was hypothesized (see Fig. \ref{Fig1}(b)).  Learning reduces the receptive field to tiny elipses representing coherent stimuli (sound waves indistinguishable for neurons due to some tolerance). Thus, neurons in the selective stratum learn individual sound waves and we observed spontaneous formation of neuronal ``clusters" (Fig. \ref{Fig3b}(b)). Neurons within a cluster detect sound waves with different phases corresponding to a single note, while rejecting the other stimuli. 

\begin{figure*}[!hbt]
\begin{center}
\includegraphics[width=0.99\textwidth]{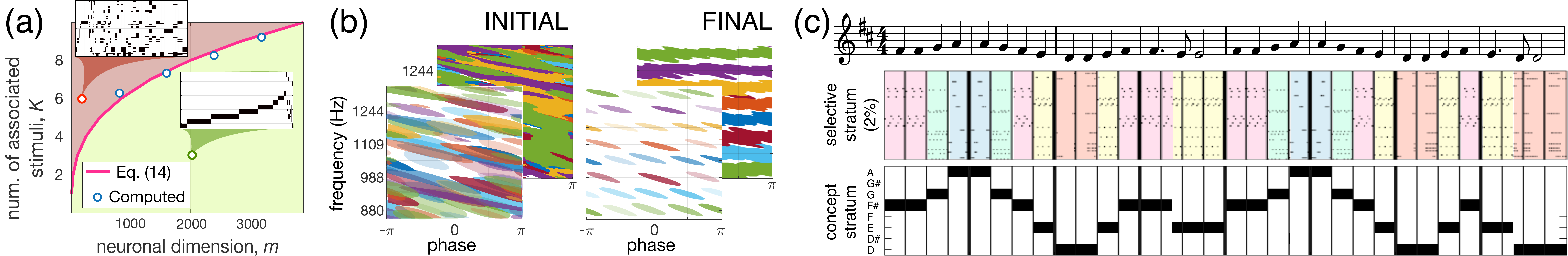}
\caption{Emergence of concept cells. (a) Working zone (shadowed by green) for association of stimuli in concepts. Insets show raster plots of the concept cells' response: a random mixture in the red zone and correct association in the green zone. (b) Formation of ``musical memory". Receptive fields in the selective and concept strata are organized into wave and note specific structures, respectively  [see also the hypothesis in Fig. \ref{Fig1}(b)]. (c) Perception of a fragment of the 9-th Symphony by Beethoven: ``Ode to joy''.}
\label{Fig4}
\end{center}
\end{figure*}

Let us now consider the second stratum composed of concept cells (Fig. \ref{Fig1}(a)). The dynamics of concept cells is also described by Eq. (\ref{eq:neuron_model}) but now as an input we use the output from the first stratum $\bfy \in \mathbb{R}_+^m$ within one time window:
\begin{equation}
\label{Stim2}
\bfs_{\rm int}(t) = \sum_{i=1}^K \bfy_i \chi_i(t), \ \ \ t\in [0,K\Delta],
\end{equation}
where $\chi_i(t)$ are overlapping rectangular time windows $[(i-1)\Delta,K\Delta)$. At the stratum output $\bfz$, we then expect to obtain codification of concepts, which are associations of $K$ individual stimuli. 

The coding is now sparse. Only a little portion of the neurons in the selective stratum responds to a single stimulus $\bfx_i$, i.e., $|\supp (\bfy_i)| \ll m$. Besides, $\bfy_i \ge 0$ and hence $\langle \bfy_j,\bfy_i\rangle \ge 0$. We also assume that after learning neurons in the first stratum are selective, i.e., $d_j=1$, $j=1,\ldots,m$. Then, $\supp (\bfy_j) \cap \supp (\bfy_i) =\emptyset$ and hence $\langle \bfy_j,\bfy_i \rangle = 0$ ($j\ne i$), which facilitates  learning.    

Repeating similar arguments for selecting the order parameter $\beta$ as provided above, after tedious calculations \cite{Suppl}, we find: 
\begin{equation}
\label{beta_concept}
\beta_{\rm cn} =  \frac{\theta_{\rm cn}\sqrt{L}\delta K \Gamma(K+\frac{1}{2})}{\theta_{\rm sl}(1-p_{\rm cn})(1-\delta) (K-1)!\sqrt{m}},
\end{equation}
where $\Gamma$ is the gamma function and $p_{\rm cn}$ is the probability that after learning, a neuron will fire to all $K$ stimuli, i.e., will become a concept cell. 

Inequality (\ref{beta_concept}) yields the following approximate condition on the neuronal dimension of CCs: 
\begin{equation}
m \propto K^3/\beta_{\rm cn}^2. 
\end{equation}
Figure \ref{Fig4}(a) shows how the number of stimuli $K$, which can be associated in a concept, scales with the neuronal dimension $m$. An increase in the association depth $K$ requires cubic increase of the input dimension $m$, which can be balanced by an increase of the order parameter $\beta_{\rm cn}$. In consequence, overloaded associations with high $K$ can result in detection of ``wrong" stimuli as being within a concept. Such an observation has been reported experimentally, when a Jennifer Aniston neuron the next day also detected Lisa Kudrow from the TV series ``Friends" \cite{QQPhys}.     

Figure \ref{Fig4}(b) illustrates the process of formation of selective and concept cells responding to musical notes. At the beginning, both selective and concept cells have messy receptive fields (see also Fig. \ref{Fig3b}(a)). Learning organizes the strata as it was advanced in Fig. \ref{Fig1}(b). Finally, concept cells are activated to particular notes regardless of the phase of sound waves. Thus, the ``brain" can now follow music. Figure \ref{Fig4}(c) illustrates the system response to a song. As expected, the selective stratum detects individual sound waves, while the concept stratum puts them together and forms the note-specific output.

In conclusion, we have shown that the emergence of concept cells is conditioned by the synaptic (input) dimension of principal neurons in feedforward connected strata. The evolution of living organisms (and ANNs) towards more complex cognitive functions requires an increase of the neuronal input dimension. A concept capable brain (or ANN) should meet the following requirements: a) At least one selective and one concept strata; b) The adequate neuronal dimensions, e.g. $n \approx 10^2$ and $m \approx 10^4$ for the selective and concept strata, respectively; c) The order parameter $\beta$ properly chosen for different strata. 

These conditions are fundamental and have no specific relations to the fine features of the model. Encephalized animals and humans satisfy constraints (a) and (b). Thus, our results support the hypothesis of a strong correlation between the level of the neural connectivity in living organisms and different cognitive behaviors such organisms can exhibit (cf. \cite{HH2012,Lobov}).  Condition (c) is related to the learning rate and hence to the magnitude of the synaptic plasticity, which differs significantly among neurons \cite{SinPlast}. It defines whether a neuron can be selective or associative. We thus suggest a hierarchy of cognitive functionality. First relay stations in the information processing, i.e., selective strata, gain extreme selectivity at intermediate dimensions ($n\approx 30-100$). The second critical transition occurs at much higher dimensions $m \approx 500-1000$ (cf. \cite{Tyukin2018}). Then, neurons located in concept strata become capable of associating multiple uncorrelated inputs of different sensory modalities into concepts. A straightforward extension of our model is an inclusion of more strata, which could encode association of primarily concepts into compound ones, as was observed experimentally \cite{QQ2}.

\textit{Acknowledgements.} This work was supported by the Russian Science Foundation (19-12-00394) and by the Spanish Ministry of Science, Innovation and Universities (FIS2017-82900P).  


\end{document}


\preprint{APS/123-QED}

\title{Supplemental material for: \\Universal principles justify the existence of concept cells}

\author{Carlos Calvo Tapia, Ivan Tyukin, Valeri A. Makarov}

\date{\today}

\maketitle

\appendix

\section{Firing probability in selective stratum at $t=0$}
\label{Corrected}
Before discussing the firing probability, let us mention useful results. 
\subsection{Hoeffding's inequality}
Let $X_i$, $i=1,\ldots,n$ be random variables with compact support $\Prb(X_i\in [a,b]) = 1$, and $\bar{X} = \frac{1}{n}\sum X_i$. Then
\begin{equation}
\label{Hoeffd}
\begin{split}
\Prb(\bar{X} - E[\bar{X}] \ge t) \le e^{-\frac{2nt^2}{(b-a)^2}}\\
\Prb(-\bar{X} + E[\bar{X}] \ge t) \le e^{-\frac{2nt^2}{(b-a)^2}}
\end{split}
\end{equation}
for some $t >0$ \cite{Hoeffding1963}. 

\subsection{Central Limit Theorem}
Let $\{X_i\}_{i=1}^{n}$ be $n$ independent random variables with zero means and standard deviations $\{\sigma_i\}_{i=1}^n$. We introduce new random variable:  
\begin{equation}
\hat{X}=\frac{\sum_{i=1}^n X_i}{\sqrt{\sum_{i=1}^n \sigma_i^2}}
\end{equation}
with the cdf $F_n(\cdot)$. Then, we have \cite{Esseen:1942}:
\begin{equation}\label{eq:Esseen}
|F_n(\hat{X})-\Phi(\hat{X})|\leq C \frac{\sum_{i=1}^n \rho_i}{\left(\sum_{i=1}^n \sigma_i^2\right)^{3/2}},
\end{equation}
where $\Phi$ is the cdf of the standard normal distribution and $\rho_i=E[|X_i|^3]$. Moreover, the constant is bounded  by \cite{Esseen:1956,Shevtsova:2010}:
\begin{equation}
0.4097\simeq \frac{\sqrt{10}+3}{6\sqrt{2\pi}}\leq C \leq 0.56.
\end{equation}

Property (\ref{eq:Esseen}), being a version of the Central Limit Theorem, implies that empirical averages of independent random variables with zero means and finite second and third moments are asymptotically normally distributed as $n\rightarrow\infty$. If no further assumptions are imposed then the convergence rate is $O(1/\sqrt{n})$.

\subsection{Decay of tails of the membrane potential}

Employing  (\ref{eq:Esseen}) and noting that $\sigma^2=E[x_i^2]E[w_i^2]=1/9$ and $\rho=E[|x_i|^3]E[|w_i|^3]=1/16$, the cumulative distribution of $v$ satisfies
\begin{equation}\label{eq:CLT_estimate}
\left|F_n\left(\sqrt{3}v\right)- \Phi\left(\sqrt{3}v\right)\right| \leq \frac{27C}{16 \sqrt{n}}\leq \frac{0.945}{\sqrt{n}}
\end{equation}
where, as above, $F_n(\cdot)$ is the corrected distribution.

Using (\ref{Hoeffd}) we get the following estimate on the firing probability to a random stimulus $\bfx$:
\begin{equation}
\Prb(v > \theta) < e^{-\theta^2/6}.
\end{equation}
Now by employing this concentration inequality together with (\ref{eq:CLT_estimate}), we find the bounds:
\begin{equation}
\label{Pt0a}
p_{\rm dw} \leq \Prb(v>\theta) \leq p_{\rm up},
\end{equation}
where
\begin{equation}
\label{Pt0b}
\begin{split}
& p_{\rm up} = \min \left\{e^{-\theta^2/6}, 1 - \Phi(\sqrt{3}\theta)+ \frac{0.945}{\sqrt{n}}\right\},\\
& p_{\rm dw} = \max \left\{0, 1 - \Phi(\sqrt{3}\theta)- \frac{0.945}{\sqrt{n}}\right\}.
\end{split}
\end{equation}

For high $n$ the bounds converge to the probability value $1 - \Phi(\sqrt{3}\theta)$, provided in the main text. We also note the exponential convergence of  $p_{\rm up}(\theta,n)$  to zero as a function of $\theta$.  This is a direct consequence of measure concentration effects.

\section{Conditions of neuronal firing in forward time: Selection of $\beta_{\rm sl}$}
We set the firing threshold small enough, e.g. $\theta = 1$. Then, with high probability, all neurons  are active, i.e., $d_j\ge 1$, and there are no lost stimuli (Fig. 2(a), main text). 

For convenience, we denote by $\bfh = \sqrt{\frac{3}{n}}\bfx_i$ the first stimulus activating the $j$-th neuron at $t^*$, i.e., $y_j(t < t^*) =0$, $y_j(t^*) >0$. Let us now find the condition that the neuron keeps ``firing'' for $t>t^*$. 

We decompose $\bfw_j$ into vectors parallel and orthogonal to $\bfh$ (by omitting the index $j$): $\bfw = \bfw_{\parallel} + \bfw_{\perp}$, where $\bfw_\parallel := q(t)\frac{\bfh}{\|\bfh\|}$ and $\langle \bfw_\parallel,\bfw_{\perp}\rangle = 0$. Then, Eq. (2c) yields:
\begin{equation}
\label{Wdyn}
\begin{split}
\dot{\bfw}_\perp =& -\alpha \| \bfh\| y q \bfw_{\perp},\\
\dot{q} =& \alpha \| \bfh\| y (\beta^2 - q^2).
\end{split}
\end{equation}
By construction, at $t=t^*$ the neuron fires, i.e., $v(t^*) = q(t^*)\|\bfh\|> \theta$. Note that $q(t\ge t^*) > 0$, otherwise $y=0$ and there is no dynamics. Selecting $\beta > \theta/\|\bfh\|$ we ensure the firing condition $y(t\ge t^*) > 0$. Then, $\bfw_\perp(t) \to 0$ and $q \to \beta$, which implies:
\begin{equation}
\label{winf1}
\lim_{t\to \infty} \bfw(t) = \beta\frac{\bfh}{\|\bfh\|}.
\end{equation}

Note that the value of $\beta$ should not be too high, since it can diminish the neuronal selectivity (see below). Choosing $\beta = \theta/\|\bfh\| + \epsilon$, where $0< \epsilon \ll 1$, ensures activity of the neuron but it requires knowledge of $\|\bfh\|$, inaccessible \textit{a priori}. Then, by using $\|\bfh\|^2 \sim \mathcal{N}(1,\frac{2}{\sqrt{5n}})$ and requiring $\Prb( \|\bfh\|^2 > \delta^2) = p_{\rm sl}$, where $\delta \in (0,1)$ is a lower bound of $\|\bfh\|$, we can set:
\begin{equation}
\label{SelBeta}
\beta_{\rm sl} =  \frac{\theta}{\delta}, \ \ \delta = \sqrt{1-\frac{2\Phi^{-1}(p_{\rm sl})}{\sqrt{5n}}}.
\end{equation}
This guarantees faring of the neuron to the stimulus $\bfh$ in forward time with a probability no smaller than $p_{\rm sl}$. Note that the higher  the neuronal dimension $n$, the higher $p_{\rm sl}$ can be chosen.

\section{Selectivity after learning}

We assume that a neuron has learnt  an arbitrary stimulus, which we denote by $\bfh\in \left \{\sqrt{\frac{3}{n}}\bfx_i \right \}$. Then, after learning $\bfw = \beta \frac{\bfh}{\|\bfh\|}$. We now estimate the probability that the neuron is silent to another arbitrary stimulus $\bfg \in \{\sqrt{\frac{3}{n}}\bfx_i\}$ ($\bfg \ne \bfh$) given $\bfh$: 
\begin{equation}
\label{P1}
\Prb(y=0  | \bfh).
\end{equation}
This can be done in several ways.

\subsection{Probability by Hoeffding's inequality}

From (\ref{P1}) we have: 
\begin{equation}
\Prb(y = 0\, | \bfh) = \Prb\left( \langle \bfh,\bfg \rangle\le \delta \|\bfh\| \, \big  | \bfh \right).
\end{equation}
By employing the Hoeffding's inequality (\ref{Hoeffd})   we get
\begin{equation}
\Prb(\langle \bfh,\bfg\rangle \ge \tau) \le e^{-n\tau^2/18}.
\end{equation}
Thus
\begin{equation}
\label{PSilent0}
\Prb(y = 0\, | \bfh) >  1 - e^{- n\|\bfh\|^2\delta^2/18}.
\end{equation}
Now, by recalling $\|\bfh\|^2 \sim \mathcal{N}(1,\frac{2}{\sqrt{5n}})$ for high enough $n$, we obtain:
\begin{equation}
\label{PSilent1}
\mathcal{P}_H = \Prb(y = 0) >   1 - e^{-\gamma(n)}.
\end{equation}
where $\gamma(n) = \frac{\delta^2n}{18}(1-\frac{\delta^2}{45})$ is an increasing function of $n$. 

\subsection{Probability by normal distribution}

By employing normal distribution, from (\ref{P1}) we get: 
\begin{equation}
\label{PSilent0}
\Prb(y=0  | \bfh)= \Phi(\delta \|\bfh\|\sqrt{n}).
\end{equation}
Then, we extend it to arbitrary $\bfh$ as above:
\begin{equation}
\label{PSilent3}
\mathcal{P}_N = \int_0^\infty \Phi(\delta\sqrt{ns}) \kappa(s;\mu,\sigma)\, \mathrm{d}s,
\end{equation}
where $\kappa(\cdot;\mu,\sigma)$ is the normal pdf with the mean $\mu = 1$ and the standard deviation $\sigma = \frac{2}{\sqrt{5n}}$. Equation (\ref{PSilent3}) corresponds to Eq. (6) in the main text.

\subsection{Comparison of two approaches}
 
The neuronal selectivity is given by [Eq. (7) in the main text]:
\begin{equation}
\label{PSilent2}
S(n,L)  = \mathcal{P}^{L-1},
\end{equation}
where $\mathcal{P}$ can be taken either from (\ref{PSilent1}) or from (\ref{PSilent3}).

\begin{figure}[!hbt]
\begin{center}
\includegraphics[width=0.48\textwidth]{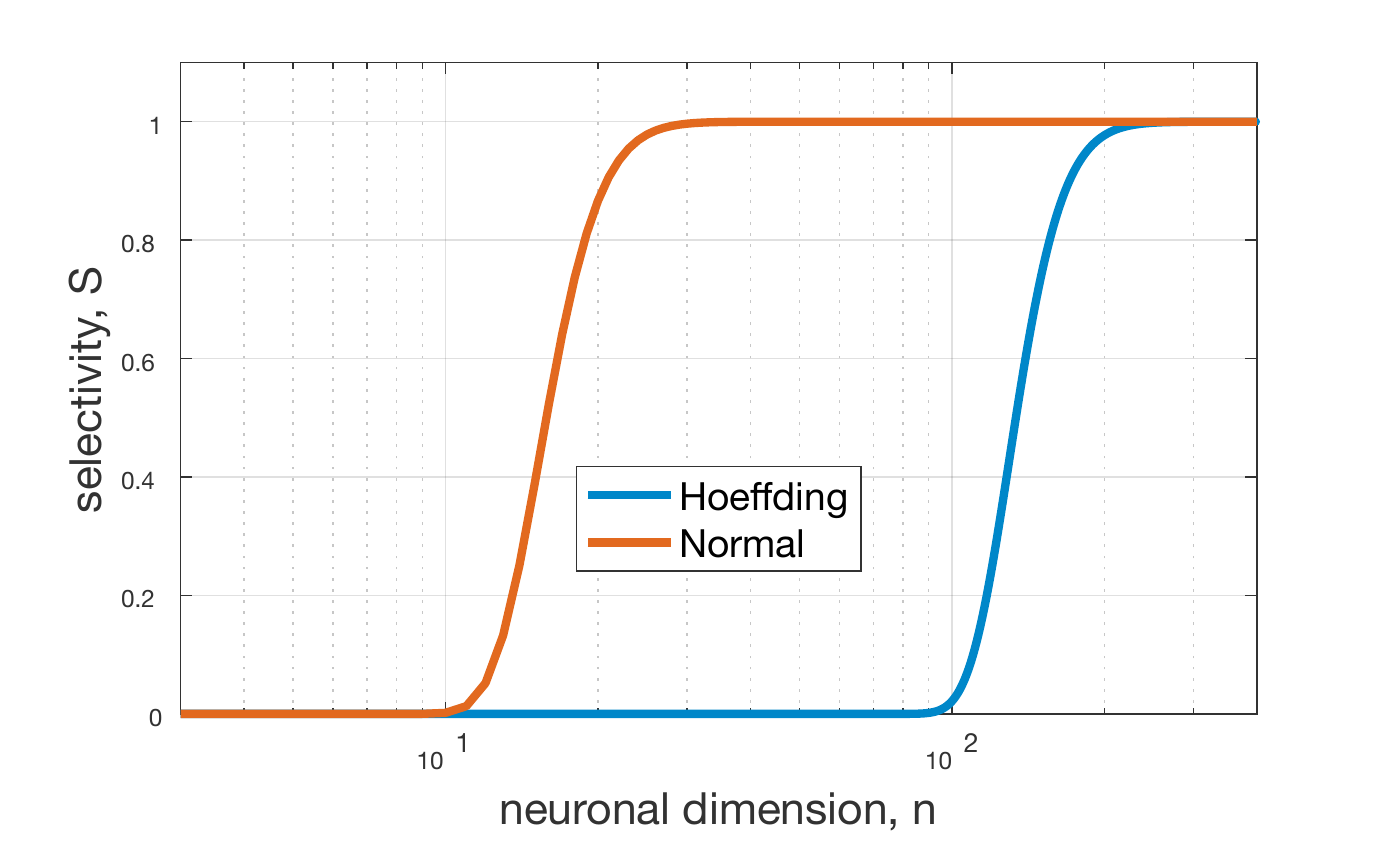}
\caption{Two estimates of $S$ (see also Fig. 3(a) in the main text, all parameter values are the same).}
\label{sFig3}
\end{center}
\end{figure}

Figure \ref{sFig3} shows the neuronal selectivity estimated by two methods. The lower bound estimated from inequalities (\ref{PSilent1}) is too conservative, while Eq. (\ref{PSilent3}) matches well the numerical results (see Fig. 3(a) in the main text).

\section{Order parameter $\beta_{\rm cn}$ for the concept stratum }

A neuron in the concept stratum receives as an input the stimulus $\bfh_k = \sum_{i=1}^k \bfy_i$, where $k$ defines the time window [see Eq. (9) in the main text]. 

\subsection{Learning condition}

At $t=0$ we assume that the neuron detects the first stimulus $\bfh_1 = \bfy_1$, i.e. $\langle \bfw(0),\bfh_1\rangle > \theta_{\rm cn}$, which is equivalent to $q(0) > \theta/\|\bfh_1\|$ in Eq. (\ref{Wdyn}). Thus, to keep firing we require
\begin{equation}
\beta_{\rm cn} > \frac{\theta_{\rm cn}}{\|\bfh_1\|}.
\end{equation}
By using Eq. (\ref{Wdyn}) we get that, at the end of the first interval $\Delta$, $\bfw \to \beta_{\rm cn} \bfh_1 /\|\bfh_1\|$. In general, the initial condition for the $k$-th interval is 
\begin{equation}
\bfw_{k0} = \lim_{t\to (k-1)\Delta} \bfw(t) \approx \beta_{\rm cn} \frac{\bfh_{k-1}}{\|\bfh_{k-1}\|}.
\end{equation}
This is equivalent to 
\begin{equation}
q_{k0} = \beta_{\rm cn} \frac{\langle \bfh_{k-1}, \bfh_{k} \rangle}{\|\bfh_{k-1}\|\|\bfh_{k}\|} .
\end{equation}

To meet the firing condition at $t = (k-1)\Delta$, we require
$q_{k0} > \theta_{\rm cn}/\|\bfh_1\|$ which yields
\begin{equation}
\beta_{\rm cn} > \frac{\theta_{\rm cn}}{\|\bfh_1\|} \frac{\|\bfh_k\|}{\|\bfh_{k-1}\|}> \frac{\theta_{\rm cn}}{\|\bfh_1\|},
\end{equation}
where we used $\langle \bfy_i,\bfy_j\rangle = 0$ for $j \ne i$ [see the main text]. Thus, given that $\alpha$ is big enough, the neuron will fire during the whole process of learning. 

Once the learning is finished, $\bfw = \beta_{\rm cn} \frac{\bfh_K}{\|\bfh_K\|}$. Then, the neuron is conceptual if
\begin{equation}
\beta_{\rm cn} > \frac{\theta_{\rm cn} \|\bfh_K\|}{\|\bfy_i\|^2}, \ \ i =1,2,\ldots, K,
\end{equation}
which is equivalent to
\begin{equation}
\beta^2  > \theta^2_{\rm cn} \frac{\sum_{i=1}^K \|\bfy_i\|^2}{\min_{i\in \{1,\ldots,K\}} \{\|\bfy_i\|^4 \}} .
\label{beta_assoc}
\end{equation}

\subsection{Estimate of $\beta_{\rm cn}$}

For convenience, let's denote:
\begin{equation}
S = \sum_{i=1}^K \|\bfy_i\|^2, \ \ \ M = \min_i \{ \|\bfy_i\|^2\}.
\end{equation}
We then set $\beta_{\rm cn} = \theta_{\rm cn} \Psi$, where $\Psi$ satisfies [Eq. (\ref{beta_assoc})]:
\begin{equation}
\label{PMS}
\Prb (M^2\Psi^2 > S) = p_{\rm cn},
\end{equation}
where $p_{\rm cn}$ is the lower probability bound. This equation ensures that the concept stratum learns at least $K$ inputs with the probability not smaller than $p_{\rm cn}$.

For further calculations, we  assume that $z:=\|\bfy\|^2$ is exponentially distributed:
\begin{equation}
f_z(z) = \left \{
\begin{array}{ll}
\lambda e^{-\lambda z}, \ \ & z >0\\
0& \mbox{ otherwise}
\end{array}
\right.
\label{fZ}
\end{equation}
Then, $S$ follows the Erlang distribution:
\begin{equation}
\label{fS}
f_S(s) = \frac{\lambda^K s^{K-1} e^{-\lambda s}}{(K-1)!}, \ \ s > 0.
\end{equation}
To find the distribution of $M$ we write:
\begin{equation}
\begin{array}{c}
F_M(m) = \Prb (\min \{z_i\} \le m) =  1 - (1-F_z(m))^K.
\end{array}
\end{equation}
where $F_z$ is the cdf of $z$. Thus,
\begin{equation}
\label{FM}
F_M(m)  = 1 - e^{-K\lambda m}, \ \ m > 0.
\end{equation}

We now can assume that $S$ and $M$ are independent and hence $f(m,s)=f_M(m)f_S(s)$. Then, Eq. (\ref{PMS}) yields 
\begin{equation}
\label{PPP}
p_{\rm cn}=\int_0^\infty f_S(s) (1-F_M(\sqrt{s}/\Psi)) \, \mathrm{d}s.
\end{equation}
By using (\ref{fS}), (\ref{FM}), (\ref{PPP}), and operating, we get
\begin{equation}
\label{Int13}
p_{\rm cn} =  \int_0^\infty \frac{u^{K-1}e^{-u - a\sqrt{u}}}{(K-1)!} \, \mathrm{d}u, \ \ a = \frac{K\sqrt{\lambda}}{\Psi}.  
\end{equation}

We now note that $a$ is a small parameter. Thus, we can approximate $e^{-u - a\sqrt{u}} \approx e^{-u}(1 - \sqrt{u}a)$ and evaluate the integral (\ref{Int13}):
\begin{equation}
p_{\rm cn} = 1 - a \frac{\Gamma (K +\frac{1}{2})}{(K-1)!}.
\end{equation}
This equation provides the estimate:
\begin{equation}
\label{1for}
\beta_{\rm cn} = \theta_{\rm cn} \sqrt{\lambda}\frac{K \Gamma (K +\frac{1}{2})}{(1-p_{\rm cn})(K-1)!}.
\end{equation}

We now note that $\lambda = 1/E[z]$ and assume that all neurons in the selective stratum have learnt stimuli, i.e.,
\begin{equation}
z = \|(\beta_{\rm sl} \|\bfh\| - \theta_{\rm sl}) \bfb\|^2,
\end{equation}
where $\bfb$ is a binary vector representing neurons activated by the stimulus $\bfh$. Thus,
\begin{equation}
E[z] = \beta_{\rm sl}^2 E[(\|\bfh\| - \delta)^2] E[\|\bfb\|^2].
\end{equation}

Then, we note that $\|\bfb\|^2 \sim \mathcal{B}(m,p)$ and hence $E[\|\bfb\|^2] = mp$. In the case that all $L$ stimuli have been learnt $p = L^{-1}$. Now we have
$E[(\|\bfh\| - \delta)^2] = 1- 2\delta E[\|\bfh\|] + \delta^2$. In the first order approximation, $E[\|\bfh\|]\approx 1$. Thus, we have
\begin{equation}
\label{Ez}
\lambda \approx \frac{L}{\beta_{\rm sl}^2 (1 - \delta)^2m}.
\end{equation}
Substituting approximation (\ref{Ez}) into Eq. (\ref{1for}) we obtain Eq. (10) in the main text.